\documentclass[pdflatex,sn-mathphys-num]{sn-jnl}

\usepackage{graphicx}%
\usepackage{multirow}%
\usepackage{amsmath,amssymb,amsfonts}%
\usepackage{amsthm}%
\usepackage{mathrsfs}%
\usepackage[title]{appendix}%
\usepackage{xcolor}%
\usepackage{textcomp}%
\usepackage{manyfoot}%
\usepackage{booktabs}%
\usepackage{algorithm}%
\usepackage{algorithmicx}%
\usepackage{algpseudocode}%
\usepackage{listings}%

\usepackage{subfig}

\theoremstyle{thmstyleone}%
\theoremstyle{thmstyletwo}%

\theoremstyle{thmstylethree}%

\begin{document}

\title[Article Title]{Good Vibes! Towards Phone-to-User Authentication Through Wristwatch Vibrations}


\author[1]{\fnm{Jakob} \sur{Dittrich}}\email{S2210455017@fhooe.at}
\author[1,2]{\fnm{Rainhard Dieter} \sur{Findling}}\email{rainhard.findling@fh-hagenberg.at}

\affil[1]{SAIL Department, University of Applied Sciences Upper Austria}
\affil[2]{\orgname{Google LLC}}

\abstract{
While mobile devices frequently require users to authenticate to prevent unauthorized access, mobile devices typically do not authenticate to their users. This leaves room for users to unwittingly interact with different mobile devices. We present GoodVibes authentication, a variant of mobile device-to-user authentication, where the user's phone authenticates to the user through their wristwatch vibrating in their pre-selected authentication vibration pattern. We implement GoodVibes authentication as an Android prototype, evaluate different authentication scenarios with 30 participants, and find users to be able to well recognize and distinguish their authentication vibration pattern from different patters, from unrelated vibrations, and from the pattern being absent.}


\maketitle

\section{Introduction}
Authentication with mobile devices is usually done from users towards devices (user-to-device authentication), using methods such as PINs, passwords, unlock patterns, or biometrics. That devices typically do not authenticate towards their users leaves room for accidentally interacting with other devices, as well as for targeted hardware phishing attacks~\cite{Findling_15_DeviceToUser}.
Device-to-user authentication~\cite{Mayrhofer_21_MobileDeviceAuthentication}, through which users can verify the authenticity of their device, is one way to reduce the risk of such errors and attacks.

In this paper we present GoodVibes authentication, which is a variant of device-to-user authentication with multiple mobile devices. When users start to interact with their phone, thereby wake the device screen, the phone authenticates to them by making the paired wristwatch vibrate in their pre-selected authentication vibration pattern.
We implement GoodVibes authentication as an Android prototype and evaluate it with 30 participants to answer the following research questions:
RQ1: How well can users recognize when a phone authenticates to them through a wristwatch vibration pattern, and when this vibration pattern is absent, incorrect, or occurs without them interacting with the phone?
RQ2: How do users perceive the usability of a phone authenticating to them through a wristwatch vibration pattern?

\section{Related Work}

Much prior work focuses on mobile user-to-device authentication through knowledge, 
tokens, or biometrics~\cite{Meng_15_SurveyingBiometricAuth,shah_19_TrensUserAuthenticationSurvey}. Such work does not address devices authenticating to users, hence leaves room for users to accidentally interact with other devices, and for targeted hardware phishing attacks~\cite{Findling_15_DeviceToUser}. 
Similar to website-based phishing, hardware phishing attacks try to trick users to interact with phishing devices that appear identical to the user owned devices. When users attempt to unlock such a device with their PIN, password, unlock pattern, or biometrics, they unwittingly disclose this authentication secret to the attackers. As attackers are already in possession of the user device from previously replacing it with the phishing device, they can immediately use that secret to access the user device and the private data on it.

Device-to-user authentication~\cite{Mayrhofer_21_MobileDeviceAuthentication} aims to make it easier for users to recognize if they would interact with another device than their own. Prior research investigates phones authenticating to their users through the phone itself vibrating in an authentication pattern when the user wakes the device screen~\cite{Findling_15_DeviceToUser}. Findings indicate that users can recognize and distinguish such vibrations (median success rate of 97.5\%). However, physical access to the user phone also gives attackers access to the vibration pattern. This allows them to configure the phishing device with the same vibration. GoodVibes authentication addresses this shortcoming, as attackers in control of the locked user phone have no access to the authentication vibration pattern.

Other vibration-based authentication includes Vibrate-to-Unlock~\cite{Saxena_11_Vibrate_to_unlock}, which uses vibration patterns on mobile phones to authenticate users to RFID tags. While this approach uses vibrations, it does so to perform user-to-device authentication, hence does not address the need for device-to-user authentication.
Prior work on website-to-user authentication investigates watermarking~\cite{Singh_11_DetectionPhising} and embedding secrets in website content~\cite{Varshney_12_PersonelSecretInformation}, which aims to make it challenging for attackers to correctly mimic a legitimate website. This requires users to visually inspect the website for the watermark or embedded secret in order to detect phishing attempts.

\section{GoodVibes Authentication}

To enroll, the user installs the GoodVibes application on their paired phone and wristwatch, and in the application selects the authentication vibration pattern in which the wristwatch should vibrate to indicate that their phone is authenticating to them.

\begin{figure}[htbp]
    \centering
    \includegraphics[width=0.6\linewidth]{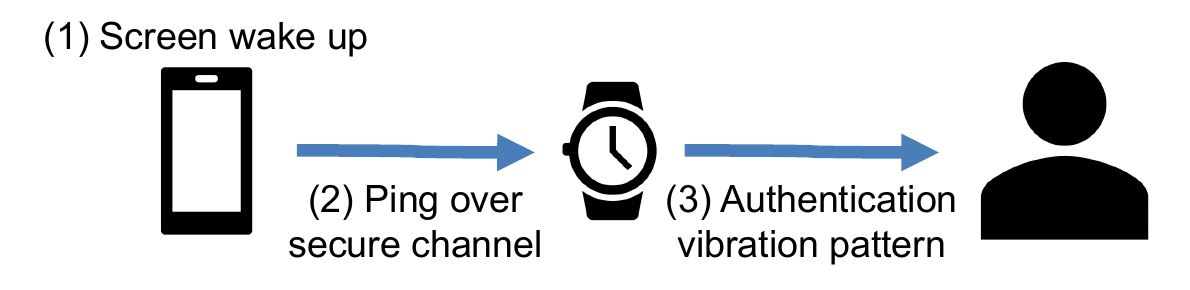}
    \caption{Overview of the GoodVibes authentication process.}
    \label{fig:setup}
\end{figure}
The GoodVibes authentication process works as follows (fig.~\ref{fig:setup}): when the user starts to interact with their phone, thereby wakes the device screen (1), the phone sends a ping to the paired wristwatch over the secure channel that exists as a result of the devices being paired (2). The wristwatch receives the ping and vibrates in the user-chosen authentication vibration pattern (3). The user feels the wristwatch vibrating in their authentication vibration pattern, which informs them that they are interacting with their phone.

If the wristwatch would not vibrate, or vibrate in a different pattern, then this would indicate to the user that they might be interacting with a phone that is not theirs. 
Also, if the user would feel their authentication vibration pattern without interacting with their phone, then this would indicate to them that someone else might be interacting with their phone.


\section{Evaluation}

We implement GoodVibes authentication as a two-part prototype app\footnote{The GoodVibes authentication prototype app is publicly available at \url{https://github.com/Jakob-Dittrich/GoodVibesAuth}.} for a paired Android phone and Android wristwatch, which can communicate securely with each other over their paired Bluetooth channel.
The prototype app can also be controlled remotely by the study supervisor to inject or suppress vibrations, to simulate and evaluate different situational authentication scenarios (sec.~\ref{sec:eval_scenarios}).
As authentication vibration patterns we use the "2" and the "1 3" patterns from~\cite{Findling_15_DeviceToUser}, with vibration bursts of 60\,ms, vibration pauses between bursts of 200\,ms, and a total duration of 180\,ms and 560\,ms, respectively.

\subsection{Evaluation Procedure}

A supervisor provided instructions and monitored the session. Participants answered demographic and technical knowledge questions. They then took a seat at a desk in an empty office room where only supervisor and participant were present, with the phone on the desk in front of them, and attached the wristwatch to their preferred hand.
50\% of participants then chose their preferred authentication vibration pattern, while the other 50\% were randomly assigned an authentication vibration pattern.
They were instructed to indicate any case of the wristwatch not vibrating in their personal authentication vibration pattern when they started to interact with the phone, and of the wristwatch vibrating in their personal authentication pattern without them having started to interact with the phone.
They were then engaged in various tasks and objectives, such as browsing a news website, playing a game, reading a page from a book, or writing a message. This simulated situations of concentration and exposure to distraction, and included points of purposefully starting and stopping to interact with the phone. 
Participants did those tasks for roughly 35 minutes, during which they were exposed to 5 different situational authentication scenarios (sec.~\ref{sec:eval_scenarios}), 24 times in total (9, 6, 3, 3, and 3 times each), in a fixed order.
They then completed a questionnaire with 5 Likert-scale questions to evaluate the usability of GoodVibes authentication.
Participants were recruited for our evaluation study through word of mouth among students of the University of Applied Sciences Upper Austria and their relatives and friends. Participants received no compensation for their participation.
30 participants (mean age 34.1 years, age std 12.3; 17 female, 13 male, 0 other) completed the study, resulting in 720 authentication results and 60 questionnaire answers.

\subsection{Evaluation Scenarios}
\label{sec:eval_scenarios}

We evaluate GoodVibes authentication in 5 scenarios that cover different cases of users possibly incorrectly identifying vibration patterns or their absence.

In scenario 1, when the user starts to interact with their phone, the wristwatch vibrates in their personal authentication vibration pattern. We measure how often users incorrectly recognize this as the vibration being absent or as not being theirs.

In scenario 2, without the user starting to interact with their phone, the wristwatch vibrates in a pattern that is not their personal authentication vibration pattern, simulating unrelated notifications, alerts, or alike. We measure how often users incorrectly recognize this as their personal authentication vibration pattern.

In scenario 3, without the user starting to interact with their phone, the wristwatch vibrates in their personal authentication vibration pattern. We measure how often users correctly identify this as their personal authentication vibration pattern.

In scenario 4, when the user starts to interact with their phone, the wristwatch does not vibrate. We measure how often users correctly identify this lack of vibration.

In scenario 5, when the user starts to interact with their phone, the wristwatch vibrates in a different pattern than their personal authentication vibration pattern. We measure how often users correctly identify this mismatch.


\section{Results and Discussion}

Overall, participants recognized vibrations and their absence well.
In scenario 1, participants incorrectly thought the vibration was absent or not their authentication vibration pattern 1\% of the time. This indicates that users can reasonably well recognize their authentication vibration pattern in normal authentication situations, and that GoodVibes authentication seems to cause a reasonably low false alert rate.

In scenario 2, participants recognized 97\% of the time that the vibration was unrelated to their personal authentication vibration pattern. Similar to the results of scenario 1, this too indicates that users can reasonably well distinguish between GoodVibes authentication vibrations and unrelated vibrations. In the remaining cases (3\%) users could easily quickly check their phone's status to ensure that it is actually secure.
This seems a sufficiently low false alert rate to not overly annoy GoodVibes authentication users, which questionnaire findings seem to confirm (see below).

In scenario 3, participants recognized 98\% of the time that the vibration was their personal authentication pattern without them having started to interact with their phone. This indicates that GoodVibes authentication would likely help users to recognize that someone else started to use their phone, e.g.\ another person mistaking the phone for theirs, or someone having stolen the phone~-- and in turn help the user to take timely countermeasures.

In scenario 4 and 5, participants recognized 91\% of the time if their wristwatch did not vibrate, and 94\% of the time if their wristwatch vibrated in a pattern that was not their personal authentication vibration pattern. Though those error rates are slightly higher than in the scenarios above, this indicates that GoodVibes authentication would likely also help users to recognize cases of them mistaking another phone as theirs, cases of hardware phishing attacks, and cases of their wristwatch coincidentally vibrating for a different reason at the same time. This indicates that GoodVibes authentication could help users to not continue their user-to-device authentication in those moments, thereby protecting their authentication secret towards their phone.

Participants with prior smartwatch experience more often correctly recognized vibrations or their absence than participants without prior smartwatch experience. Over all scenarios combined, 9 participants who use a smartwatch daily achieved 97\% correct vibration recognition, and 8 participants who had sometimes used a smartwatch achieved 99\%~-- while 13 participants without prior smartwatch experience achieved 89\%.
This indicates that users seem able to learn to recognize and distinguish wristwatch vibrations, which in turn indicates that GoodVibes authentication users without prior smartwatch experience should be able to over time learn to distinguish their authentication vibration pattern from other vibrations.
Also, over all scenarios combined, participants who could choose their personal authentication vibration pattern achieved better results over participants who were assigned a pattern (2\% and 5\% overall error rate, respectively). This indicates that it is beneficial to let users choose which vibration pattern they prefer to use with their GoodVibes authentication.

Questionnaire results indicate that participants found GoodVibes authentication easy to use (mean 4.9, std 0.3, from 1 "very difficult" to 5 "very easy"), fast to use (mean 5, std 0, from 1 "very slow" to 5 "very fast"), and easy to adapt to (mean 4.9, std 0.3, from 1 "very hard" to 5 "very easy"). Slightly more users considered it likely than unlikely that they would use GoodVibes authentication if it were available today (mean 3.4, std 0.9, from 1 "very unlikely" to 5 "very likely").

\medskip

In summary we can answer the posed research questions as follows:
RQ1: users seem to well recognize wristwatch vibration patterns through which their phone would authenticate to them (99\%), and also to well distinguish them from unrelated vibrations (97\%). Users furthermore seem to recognize well if their authentication vibration pattern occurs without them interacting with their phone (98\%), as well as when they expect the pattern to occur but it instead is absent (91\%) or different (94\%).
RQ2: users seem to find their phone authenticating to them through wristwatch vibrations easy to use (4.9/5), fast to use (5/5), easy to adapt to (4.9/5), and slightly more users considered it  likely than unlikely that they would use such an approach (3.4/5).


\section*{Conclusion}We presented GoodVibes authentication, where phones authenticate to their users through vibrations of their paired wristwatch. We evaluated how well users can recognize and distinguish such authentication vibration patterns, or their absence, in 5 different authentication scenarios, and how users perceive the usability of such authentication vibrations.
Findings include that users seem to be able to well recognize and distinguish their authentication vibration pattern from other vibrations, their pattern occurring without them interacting with their phone, as well as the pattern being absent or different. Users seem to perceive the usability of their phone authenticating to them over vibrations of their wristwatch as good.

Our work is limited by accounting only for a limited threat model without sophisticated attackers, and by assessing vibration pattern recognizably and usability with users who only used GoodVibes authentication for a short time in a lab study environment. Future research could expand our work by considering a wider threat model with different attackers, and by assessing outside-lab vibration pattern recognizably and usability in a long-term real-world study where participants are exposed to diverse distractions and environmental factors that occur in their everyday lives.


%
\bibliography{sn-bib-short}

\end{document}